\documentclass[12pt,preprint]{aastex}

\shorttitle{Photometric Redshift}
\shortauthors{Wiklind}

\begin{document}

\title{Photometric redshift using the far-infrared SED}

\author{T. Wiklind}
\affil{European Space Agency Space Telescope Division\\Space Telescope Science
Institute\\3700 San Martin Dr., Baltimore MD 21218, USA}
\email{wiklind@stsci.edu}

\begin{abstract}
A method for using the sub-millimeter band to determine photometric redshifts
of luminous high-z dusty galaxies is presented. It is based on the observation that
local ultra-luminous IR galaxies show an average spectral energy distribution
which has a remarkably small dispersion for wavelengths $\lambda > 50\mu$m.
The shape of the long wavelength part is independent on whether the local
ULIRG possesses an AGN or not. 
A local template is derived from a sample of ULIRG data presented by Klaas
et al. (2001). Using this template and the internal dispersion in its shape,
it is shown that observations at two fixed submm wavelengths, e.g. 450$\mu$m
and 850$\mu$m, can be used to determine the photometric redshift of galaxies
in the redshift range $1 < z < 5$. The uncertainty in the redshift arising
from the template SED amounts to $\Delta z/z \sim 0.3$, with an improvement
at the higher redshift range.
Implications for using this method together with the large instantaneous bandwidth
of the Atacama Large Millimeter Array (ALMA) for deriving spectroscopic
redshifts using CO emission is discussed.

\end{abstract}

\keywords{techniques: photometric, galaxies: distances and redshifts, galaxies: high redshift,
galaxies: photometry, submillimeter}

\section{Introduction} \label{intro}

The spectral shape of the far-infrared background suggests that as much as
half of the energy ever emitted by stars and AGNs has been absorbed by dust
grains and re-radiated in the far-infrared (Puget et al. 1996; Fixsen et al.
1998; Gispert, Lagache \& Puget 2000).
Recent observations at sub-millimeter wavelengths have revealed a population
of objects believed to be dusty high redshift far-infrared luminous galaxies
(Smail et al. 1997; Barger et al. 1998; Hughes et al. 1998; Eales et al.
1999).
These sources emit most of their bolometric luminosity in the far-infrared
(FIR), and are believed to account for a large portion of the far-infrared
background (e.g. Blain 1999).
Although detailed redshift information is still lacking for most of these
sources, they are believed to be situated at $z > 1$. Number-counts then
indicate a large excess of sources compared to no-evolution models based
on optical surveys (Guiderdoni et al. 1998; Blain 1999; Scott et al. 2002b),
suggesting that the optically derived star formation density is
significantly underestimated at high redshifts.
This strong evolution is difficult to reproduce in semi-analytical
hierarchical Cold Dark Matter models (Devriendt \& Guiderdoni 2000;
Kaviani, Haehnelt \& Kauffmann 2002). Assuming similar dust and FIR
properties between the high- and low-z objects, the CDM models
under-estimate the observed number counts by almost two orders of
magnitude (Kaviani et al. 2002).

A remaining and central uncertainty in the discussion about the
properties of the dusty high redshift population is their actual
redshift distribution.
Despite detection of $>$100 sub-millimeter sources, only a handful have
secure optical or near-infrared counterparts (cf. Smail et al. 2002). This
lack of counterparts shows that the submm sources are very faint (R$\geq$25
and K$\geq$21), making it difficult to obtain redshifts even with 8-10m
class telescopes, should the counterparts be identified. The presence of
large amounts of dust also makes optical/NIR photometric redshift
determinations unreliable, should the continuum be detected.

Ultra-deep long wavelength radio continuum surveys have shown that many of
the submm detected objects are related to weak radio sources (Smail et al.
2000). Chapman et al. (2002b) find that $\sim$2/3 of sources detected at
850$\mu$m are also radio-identified at sub-mJy levels.
For local galaxies there exists a very tight correlation between the
long wavelength radio continuum and the far-infrared at 60$\mu$m and 100$\mu$m,
extending over almost four orders of magnitude in luminosity (Condon 1992).
The physical mechanism for this correlation is believed to be the formation
of massive stars, which heats the dust through their UV radiation, producing
the FIR emission, and supernova explosions producing relativistic electrons
which give synchrotron emission in the radio regime.
Based on this tight correlation between non-thermal radio and thermal
far-infrared fluxes for local galaxies, Carilli \& Yun (1999) devised a
technique to use the radio-FIR spectral index as a redshift indicator for
distant submm sources. The technique uses the fact that for a fixed observed
wavelength, the flux of synchrotron emission decreases with increasing
redshift, while the opposite is true for the Rayleigh-Jeans part of the dust
continuum. The first implementation used two local starburst galaxies as
templates (see also Barger et al. 1998). A follow-up study (Carilli \& Yun
2000) used 17 low-z star-forming galaxies as templates. The result is a
submm-radio spectral index which depends on the redshift of the source.

There are several caveats with this radio-FIR relation. The spectral
index depends not only on the redshift of the source, but also on several other
parameters, most notably the dust temperature, and the radio continuum spectral
index, for instance non-thermal vs. thermal contributions and the possible
presence of an AGN.
The decrease of the observed radio continuum with increasing redshift makes
it increasingly demanding to detect the sources as $z$ increases.
The radio-FIR relation is therefore practical up to $z\sim 2$.
Dunne et al. (2001) discuss the various dependencies,
comparing with a local sample of FIR bright galaxies.
Nevertheless, although the radio-FIR technique is too crude to be used as
a photometric redshift indicator for individual sources, becomes unreliable
at $z>2$, and care has to be exercised in order not to include radio-loud AGNs,
it has proven that the submm detected sources, seen as a class, are indeed
situated at high redshifts.
Very little is known about these sources from other wavelength bands, therefore
this information has been crucial for establishing the presence of a high
redshift population of dusty, luminous and, in the few known cases,
massive galaxies.

\medskip

Several authors have speculated on the possibility of using two or more
wavelength bands in the sub-millimeter as a mean of deriving photometric
redshifts (e.g. Hughes et al. 1998; Blain 1999; Blain, Barnard \& Chapman
2002). The conclusion has been that this is unfeasable due to a degeneracy
between the redshift and dust temperature (see Sect.~\ref{degeneracy}).
In a recent study by Hughes et al. (2002) the submm photometric method
is revived using local starburst galaxies and a statistical treatment.

In this paper an alternative method for determining the redshift of submm
detected sources is presented, using only the Rayleigh-Jeans part of the
dust SED. It is based on the existence of a template SED for local ULIRGs,
which shows a remarkably uniform shape at wavelengths
$60\mu$m$ < \lambda < 1250\mu$m. Assuming that ultraluminous IR galaxies
at high redshift have similar properties, the flux ratio of two submm
bands, for instance at 850$\mu$m and 450$\mu$m, is thus sensitive to the
redshift of the source in the range $1 < z < 5$. A comparison with submm
flux ratios for a sample of high-z AGNs (mostly gravitationally lensed)
shows that this technique is accurate enough to allow a redshift
determination to within $\sim$30\%.
With the advent of new submm telescopes and instruments, such as APEX
(Atacama Pathfinder EXperiment), ALMA (Atacama Large Millimeter Array)
and SCUBA II, some of them capable of operating up to 325$\mu$m, the use
of submm flux ratios is foreseen to become a technique to quickly estimate
approximate redshifts of large numbers of submm detected sources.

\section{Dust Spectral Energy Distributions} \label{dustsed}

In this section I will briefly discuss dust spectral energy distributions
and how these can be used to describe physical properties of unresolved
dust sources observed in the IR-to-submm. The main conclusion is that
extreme care has to be exercised when doing this interpretation since
it can easily lead to wrong inferences about dust mass and the physical
size of the emitting region.

\subsection{Models} \label{models}

The far-infrared spectral energy distribution of dust emission is
purely thermal and usually represented by a modified blackbody curve,
$S_{\nu} \propto \nu^{\beta} B_{\nu}(T_{\rm d})$, where $B_{\nu}$
is the Planck function, $T_{\rm d}$ the dust temperature and
$\nu^{\beta}$ is the grain emissivity power-law index.
The index is in the range $\beta = 1 - 2$. A single-size and
uniform grain composition would lead to $\beta = 1$. When $\beta > 1$
it is indicative of a distribution of grain sizes and/or compositions.
Such simple representation has successfully been used for cold dust
components where a large part of the SED is optically thin. When
$\tau \approx 1$ or larger, the observed dust emission needs to be
described by an expression of the form:
\begin{eqnarray}
S_{\nu} & = & \Omega_{s} B_{\nu}(T_{\rm d}) \left(1-e^{-\tau_{\nu}}\right)\ ,
\label{eq1}
\end{eqnarray}
where $\Omega_{s}$ is the solid angle of the source emissivity distribution
and $\tau_{\nu}$ is the opacity of the dust.
Setting $\tau_{\nu} = \left(\nu/\nu_0\right)^{\beta}$ gives
$S_{\nu} \propto \nu^{\beta}B_{\nu}(T_{\rm d})$ for $\tau_{\nu} \ll 1$
and $S_{\nu} \propto B_{\nu}(T_{\rm d})$ for $\tau_{\nu} \gg 1$.
The critical frequency $\nu_0$ is the frequency where $\tau_{\nu}=1$.
The FIR luminosity is obtained by integrating Eq.~\ref{eq1} over all frequencies:
\begin{eqnarray} \label{firlum}
L_{\rm IR} & = & 4\pi \left(1+z\right)^{3} D_{\rm A}^{2}
\int\limits^\infty_0{S_{\nu}\,d\nu}\ ,
\end{eqnarray}
where $D_{\rm A}$ is the angular size distance\footnote{Unless otherwise stated, all
distances will be expressed in Mpc, frequencies in GHz and fluxes in Jansky.}
With the substitution $x = h\nu/kT$, we can express Eq.~\ref{firlum} in a form
directly accessible for integration (e.g. Wiklind \& Alloin 2003):
\begin{eqnarray} \label{firint}
{{L_{\rm IR}} \over {{\rm L_{\odot}}}} & = & 
8.53 \times 10^{10}\,\left(1+z\right)^{3}\,D_{\rm A}^{2}\,
T_{\rm d}^{4}\,\Omega_{s}\,
\int\limits^\infty_0{{{x^3 \left(1 - e^{-\left(a x\right)^{\beta}}\right)} \over
{e^{x}-1}}\,dx}\ ,
\end{eqnarray}
The integral can be evaluated numerically with appropriate values of
the parameter $a = kT_{\rm d}/h\nu_0$.
The solid angle $\Omega_s$ is a parameter derived in the fitting procedure.
In the event of a single dust component and a source unresolved by the
telescope, $\Omega_s$ can be estimated from the measured flux $S_{\nu_{\rm r}}$ at
a given restframe frequency $\nu_r$
\begin{eqnarray} \label{omega}
\Omega_s & = & {{S_{\nu_r}} \over
{B_{\nu_r}(T_{\rm d})\left(1-e^{-\left(\nu_r/\nu_0\right)^{\beta}}\right)}}
\nonumber \\
 & \approx & 
6.782 \times 10^{-4}\,{\nu_r}^{-3}\,
{S_{\nu_r}}\,
\left[{{e^{h\nu_r/kT_{\rm d}}-1} \over 
{1-e^{-\left(\nu_r/\nu_0\right)^{\beta}}}}\right]\ .
\end{eqnarray}
A rough estimate of the equivalent spherical radius of the emitting region
can be obtained from $\Omega_s$; $r \approx D_{\rm A}\sqrt{\Omega_s/\pi}$.

\medskip

An estimate of the dust mass from the infrared flux requires either
optically thin emission combined with a knowledge of the grain properties,
or optically thick emission and a knowledge of the geometry of the
emission region (cf. Hildebrand 1983).
The grain properties are characterized through the macroscopic mass
absorption coefficient, $\kappa_{\nu}$ (cf. Hughes, Dunlop \& Rawlins
1997).
Combining several estimates (cf, Hildebrand 1983; Hughes et al. 1997),
the mass absorption coefficient can be described as:
\begin{eqnarray}
\kappa_{\nu_r} & \approx 0.15\,\left({{\nu_r} \over
{375\,{\rm GHz}}}\right)^{1.5}\ {\rm m^{2}\ kg^{-1}}\ .
\end{eqnarray}
This expression corresponds to a grain composition similar to that
found in the Milky Way. At frequencies where the emission is
optically thin, the dust mass can now be determined from
\begin{eqnarray} \label{mdust}
M_{\rm d} & = &
{S_{\nu_{obs}} \over {\kappa_{\nu_r} B_{\nu_r}(T_{\rm d})}}
D_{\rm A}^{2}(1+z)^{3}\ ,
\end{eqnarray}
which can be expressed as:
\begin{eqnarray}
M_{\rm d} & \approx &
4.08 \times 10^{4}\,S_{\nu_{obs}}\,D_{\rm A}^{2}
\left({{\nu_r} \over {375\,{\rm GHz}}}\right)^{-9/2}
\left(e^{h\nu_r/kT_{\rm d}}-1\right)(1+z)^{3}\ \ 
{\rm M_{\odot}}\ .
\end{eqnarray}

\subsection{Multiple components}

In reality, it is likely that a galaxy contains several different
dust components each characterized by a different set of parameters.
Lacking observational data with sufficient spatial resolution,
it is usually assumed that the dust emission can be decomposed
into three main components (e.g. Rowan-Robinson 1986, 1992);
cool dust associated with a cirrus component heated by the diffuse
interstellar radiation field, warm dust associated with regions of
massive star formation, and, if the galaxy harbors an AGN, a hot
component representing dust heated by the central AGN. These three dust
components will dominate the submm, FIR and mid-IR wavelength regimes,
respectively. There are, however, no assumptions about their actual sizes
or their geometries.

More elaborate models solve the radiation of transfer in a dusty medium
containing heating sources (e.g. Granato, Danese \& Franceschini 1996;
Efstathiou, Rowan-Robinson \& Siebenmorgen 2000).
Due to the large number of parameters, involving dust variables as a
function of the geometry, these methods can only be safely applied
to sources and parts of the SED which are known to be associated with
a single heating source, e.g. an AGN.

\medskip

The representation of the dust SED discussed in Sect.~\ref{models} involves
four parameters per component. Since most observed SEDs contain only a
few data points, a fit of Eq.~\ref{eq1} is prone to a relatively large
degree of freedom, especially if more than one dust component is used.
Moreover, at long wavelengths there exists a degeneracy between the dust
temperature $T_{\rm d}$ and the grain emissivity power-law index $\beta$
(e.g. Blain et al. 2002). This degeneracy is difficult to resolve without
an observationally well sampled SED.

\begin{figure}
\plotone{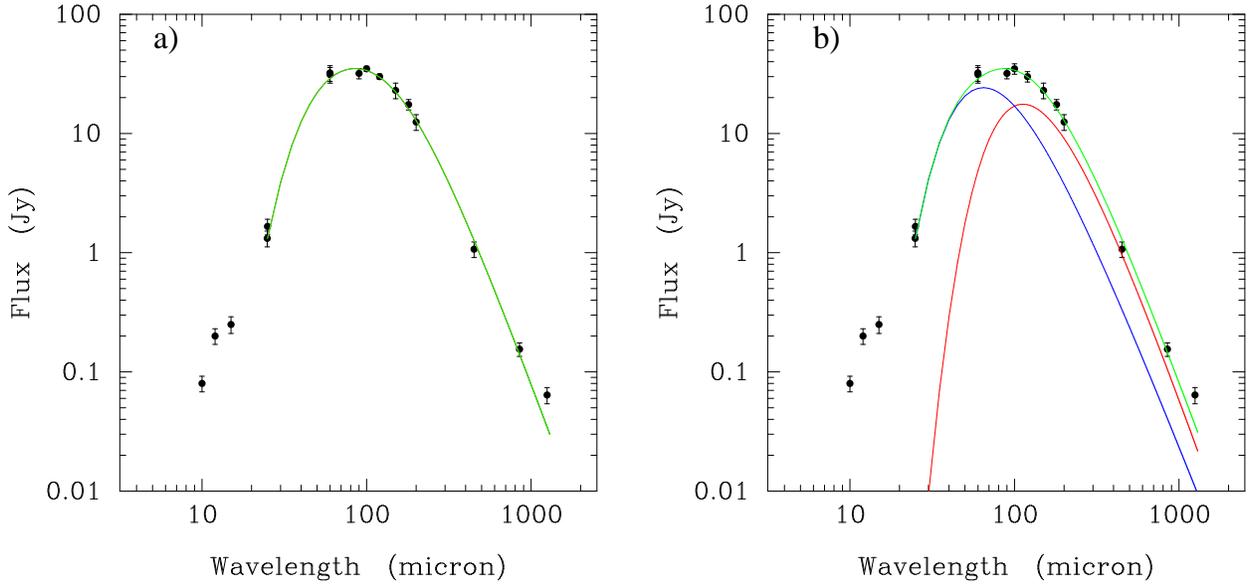}
\caption{{\bf a)}\ Fit of a one-component modified balckbody function to the
observed flux densities of the local ULIRG 17208--0014. The parameters for the
fit are: $\Omega = (9.1 \pm 1.1) \times 10^{-14}$, $\beta = 1.8 \pm 0.2$, 
$\nu_0 = (1.4 \pm 0.2) \times 10^{12}$ Hz, T$_{\rm d} = 59 \pm 1.4$ K.
{\bf b)}\ Fit of a two-component modified blackbody function to the same
flux density values. The fit is as good as the one-component and results in
the following values: 
$\Omega = 34.4 \times 10^{-14}$, $\beta = 1.6$, 
$\nu_0 = 8.6 \times 10^{12}$ Hz, T$_{\rm d} = 53$ K, for the
warm component, and 
$\Omega = 371 \times 10^{-14}$, $\beta = 2.0$, 
$\nu_0 = 5.9 \times 10^{12}$ Hz, T$_{\rm d} = 27$ K, for the
cold component.
\label{dustseds}}
\end{figure}

Due to the observationally sparsely sampled dust SEDs, the dust
emission is often characterized by fitting a single component.
This usually gives a reasonable fit to the observed data points,
but should not be over-interpreted in terms of physical conditions
of the dust grains or of global dust parameters, especially dust mass.
This is illustrated in Fig.~\ref{dustseds} where the dust SED of
the local ULIRG 17208-0014 has been fitted using both one and two dust
component SEDs of the type described above (see also Klaas et al. 2001).
The single component SED
is characterized by a dust temperature of 59 K and an exponent
$\beta = 1.8$. The two component SED has warm dust grains characterized
by a temperature of 53 K and $\beta = 1.6$, and a cool component with
T$_{\rm d} = 27$ K and $\beta = 2.0$. While the two component SED has
critical frequencies $\nu_0$ corresponding to 35$\mu$m and 51$\mu$m,
respectively, the single component SED becomes optically thick already
at 214$\mu$m.
Both models give approximately the same L$_{\rm FIR}$, showing that
they describe the {\it shape} of the SED reasonably well. The single
component SED, however, gives a total dust mass which is 3.5 times 
smaller than the two-component SED. This is due to the highly non-linear
relation between M$_{\rm dust}$/L$_{\rm FIR}$ as a function of dust
temperature; while M$_{\rm d} \propto \exp{(h\,\nu_{r}/k\,T_{\rm d})}$,
(approximately $\propto T_{\rm d}^{-1}$ at long wavelengths), the
luminosity L$_{\rm FIR} \propto T_{\rm d}^{4-5}$. It is thus possible
to `hide' a large amount of dust in a low temperature component. 
In terms of typical size scales, the cool dust extends over an area
which is $\sim$150 times larger than the warm component.

Although the resulting $\chi^2$ values for the fits shown in
Fig.~\ref{dustseds} are similar, they describe dust components with
widely different physical properties. The two fits, however, both
give a satisfactory description of the overall {\it shape} of the
SED. This is important to keep in mind when discussing and comparing
dust properties of different types of objects, and, as argued here,
when constructing a template SED.

\subsection{Degeneracy between T$_{\rm d}$ and redshift} \label{degeneracy}

Lowering the temperature of a modified blackbody curve is equivalent
to shifting the entire spectrum towards longer wavelengths. A similar
shift can be achieved by keeping the temperature constant but moving
the source to a larger redshift. Hence, without knowing the dust
temperature apriori, it is impossible to distinguish between a source
at large redshift with a given T$_{\rm d}$ and a source at low redshift
with a dust SED characterized by a lower temperature (e.g. Blain 1999;
Eales et al. 1999; Blain, Barnard \& Chapman 2002).

In order to use two flux densities at the FIR/submm part of a dust
SED to determine a photometric redshift of the source, it is thus
necessary to know the shape of the SED in sufficient detail. This
can be done if the global physical parameters for the emission can
be established. If this is not the case, the degeneracy between the
dust temperature and redshift makes progress essentially impossible.

\begin{figure}
\plotone{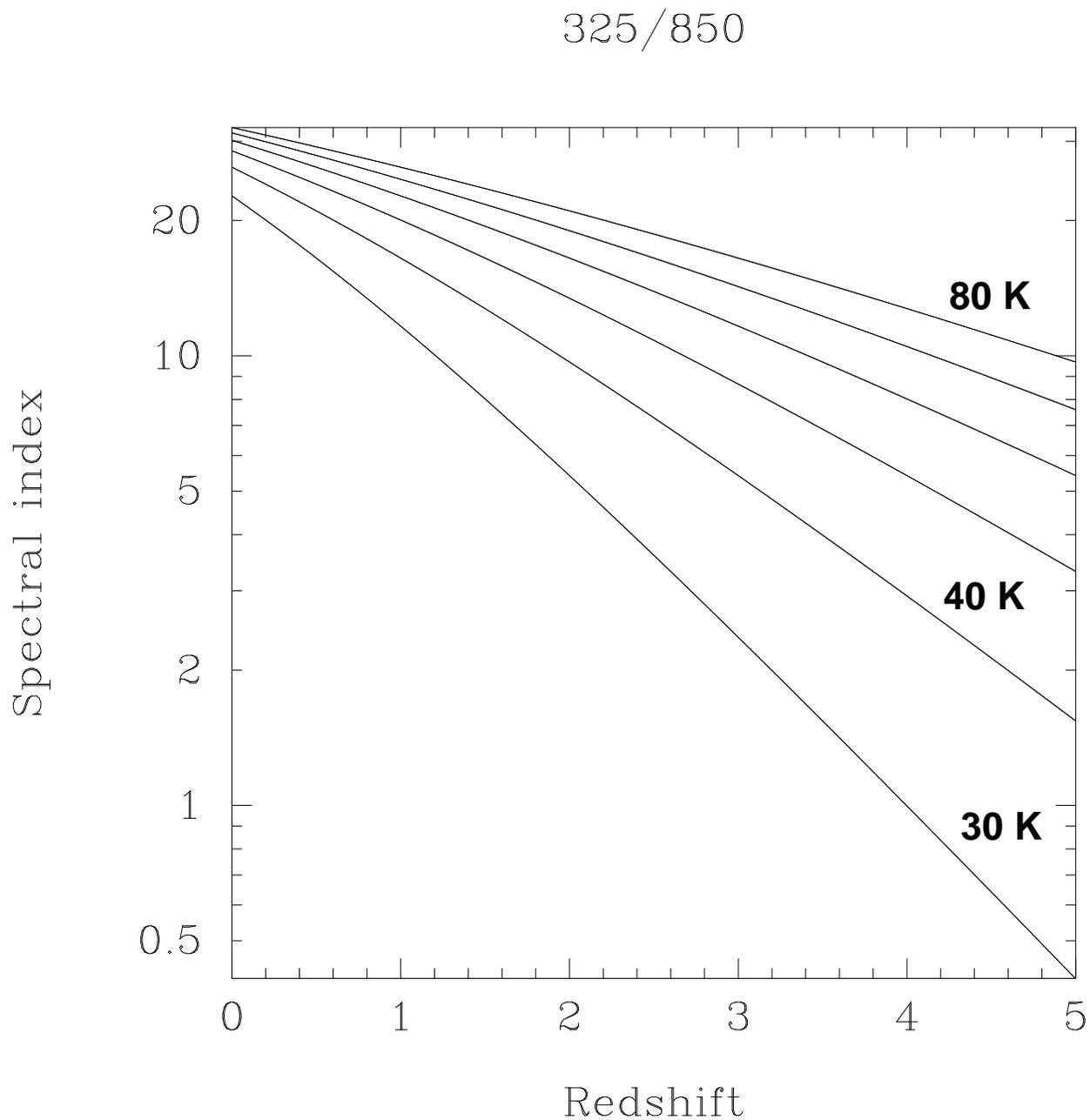}
\caption{The flux ratio at fixed observed wavlengths 325$\mu$m/850$\mu$m
for a modified blackbody spectral energy distribution as a function of
redshift of the source. The flux ratios for 6 different dust temperatures
are shown, illustrating the degeneracy between redshift and dust temperature
for a given flux density ratio.
The dust SED is represented by:
$f_{\nu} \propto \nu^{\beta}\,B_{\nu}({\rm T}_d)$, where $\beta = 1.8$.
\label{specind}}
\end{figure}

The effect of this degeneracy is illustrated in Fig.~\ref{specind} for
a simple dust SED. The ratio of flux densities at 325$\mu$m and 850$\mu$m
of a single component modified blackbody function (given by Eq~\ref{eq1})
are plotted as a function of redshift. The result for six different dust
temperatures between 30 K and 80 K are shown. Despite the fact that
the gradient of the flux ratio as a function of redshift is quite steep,
the effect of the dust temperature precludes its use as a redshift
indicator unless T$_{\rm d}$ is known.

\medskip

Attempts to characterize local FIR luminous galaxies in terms of
dust parameters show a relatively broad range in T$_{\rm d}$ and
$\beta$ (e.g. Sanders \& Mirabel 1996; Klaas et al. 2001).
As shown above, a sparsely sampled dust SED can
be equally well modeled with one and two component dust SEDs,
with different sets of dust parameters, leading to an even larger
spread in parameter values for different sources. Similarly good
fits can also be achieved for different dust temperatures by varying
the index $\beta$.
The {\it shape}, however, remains basically identical and the ratio
of FIR/submm flux densities are approximately the same for the various
fits.

\section{Photometric Redshift} \label{analysis}

\subsection{The local template} \label{template}

Local ULIRGs are characterized by a dominating FIR luminosity, 
L$_{\rm FIR} \geq 10^{12}$ L$_{\odot}$, large dust masses, large
molecular gas masses and signs of strong gravitational interaction
and/or merging (e.g. Sanders \& Mirabel 1996; Rigopoulou et al. 1999).
In addition, the dust and gas is strongly concentrated to the central
regions of the galaxies (Downes \& Solomon 1998). The heating source
of the large FIR luminosity is in many cases pure star formation, but
a significant fraction of all ULIRGs also contain an AGN (Rigopoulou
et al. 1999; Klaas et al. 2001).
Local ULIRGs have properties which in many respects are similar to
those of the submm detected population, especially those few that
have known redshift and have been studied in greater detail (cf.
Ivison 2000). Two submm detected galaxies are known to contain large
amount of molecular gas as seen through their CO emission (Frayer
et al. 1998, 1999). Using the standard conversion ratio between
CO and molecular hydrogen, H$_2$, the gas masses are in the range
$(0.5-2.0) \times 10^{11}$ M$_{\odot}$, showing that these objects are
indeed as massive and gas-rich as local ULIRGs.
Local ULIRGs are therefore the most likely low-z counterparts to the
high redshift dusty submm objects.

A recent compilation of NIR-to-submm observations of 41 local
ultraluminous infrared galaxies was presented in Klaas
et al. (2001).
The sample consists of galaxies with L$_{\rm FIR} > 10^{12}$ L$_{\odot}$,
with most objects having redshift $z \leq 0.1$.
The observational data was obtained with several different instruments.
The NIR bands (1.2 and 2.2$\mu$m) was observed using the 2.2m Calar
Alto telescope. The IR data between 10-200$\mu$m was obtained using the
ISO. Sub-millimeter data at 450$\mu$m and 850$\mu$m was obtained for a
sub-set of 16 objects using the JCMT. A few millimeter continuum observations
were also done with the SEST (Swedish-ESO Submillimeter Telescope).
The SEDs represent the most complete set of IR/FIR/submm photometric data
obtained for local ULIRGs to date.
Given the observed similarities between submm detected sources and the
local ULIRGs, the Klaas et al. sample is well suited for creating a
`template' SED to be used for the high redshift counterparts.

\medskip

In October 2002 we observed 8 of the sources in the Klaas et al. (2001)
sample using the SIMBA bolometer array at the Swedish-ESO-Submillimeter
telescope (SEST) in Chile. The SIMBA instrument is a 37-channel bolometer
array operating at 1250$\mu$m (see Nyman et al. 2001 for a description
of SIMBA). It uses a fast mapping technique and achieves a sensitivity
of $\sim$70\,mJy\,s$^{-1/2}$\,pixel$^{-1}$. Four of the local ULIRGs
were positively detected. The results are presented in Table~\ref{sesttable}.
The 1250$\mu$m result for the source 17208-0014 is used in
Fig.~\ref{dustseds}, where the long wavelength data implies the presence
of a cold dust component. This aspect of the dust continuum will be
discussed in a forthcoming paper. The 1250$\mu$m data is not used in
the present analysis.

\medskip

\begin{figure}
\plotone{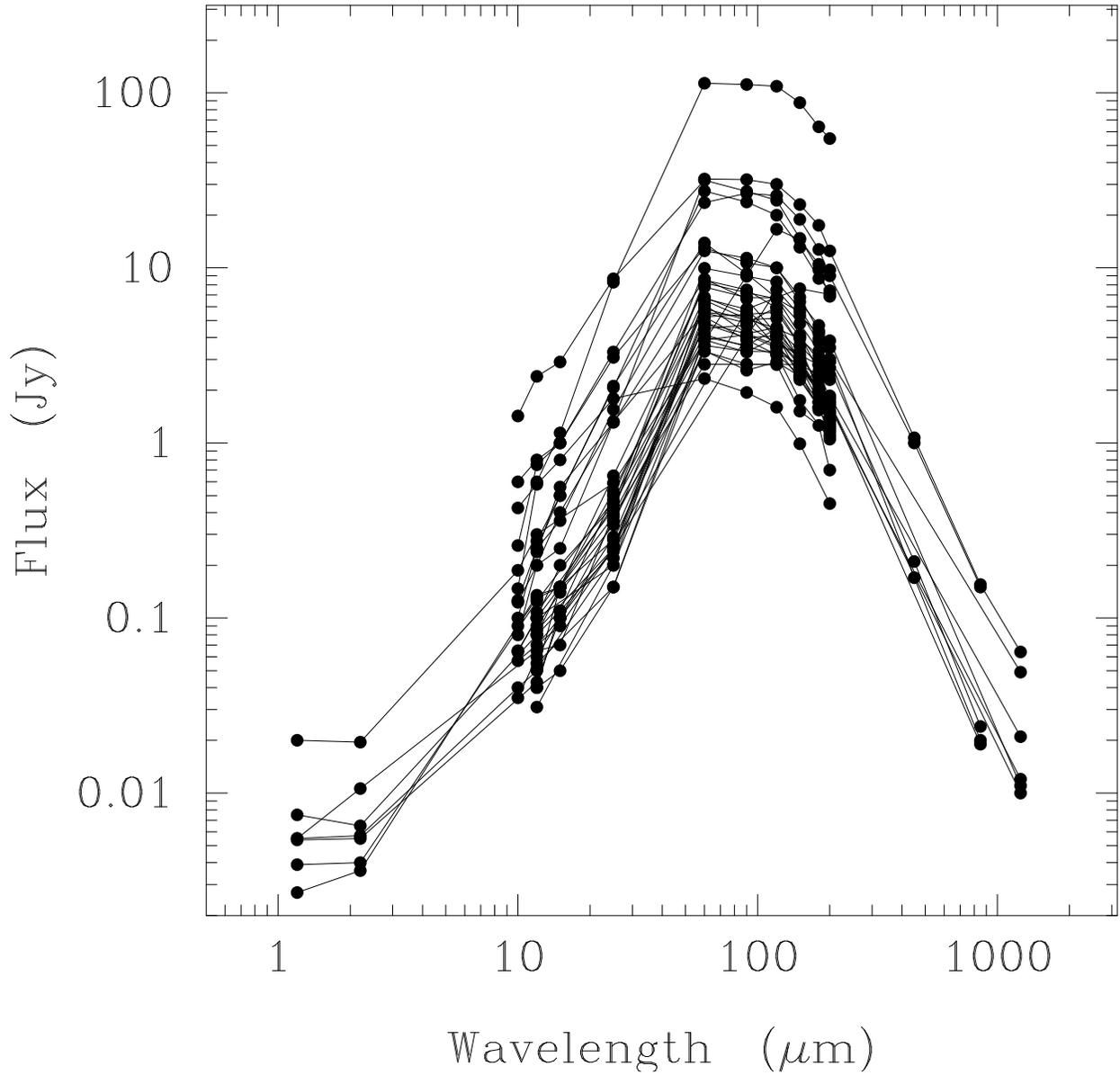}
\caption{The spectral energy distributions of the 37 ULIRGs part of
the local sample (from Klaas et al. 2001). The observed fluxes are 
plotted without any corrections or normalization. Despite a range
of fluxes over two orders of magnitude, the general shape of the
SEDs remains the same.
\label{sedplot1}}
\end{figure}

In order to avoid redshift corrections to the local sample, 4 objects
from the Klaas et al. (2001) source list, with redshifts in the range
$z=0.2-0.3$, were discarded. The remaining 37 objects all have
$z \leq 0.1$. The SEDs of these 37 sources, as tabulated by Klaas
et al., are plotted in Fig.~\ref{sedplot1}. No corrections or
normalization have been done to the SEDs at this stage. It is clear
from the figure that although the observed flux densities vary by
two orders of magnitude, the general shapes of the SEDs appear quite
uniform.

\begin{figure}
\plotone{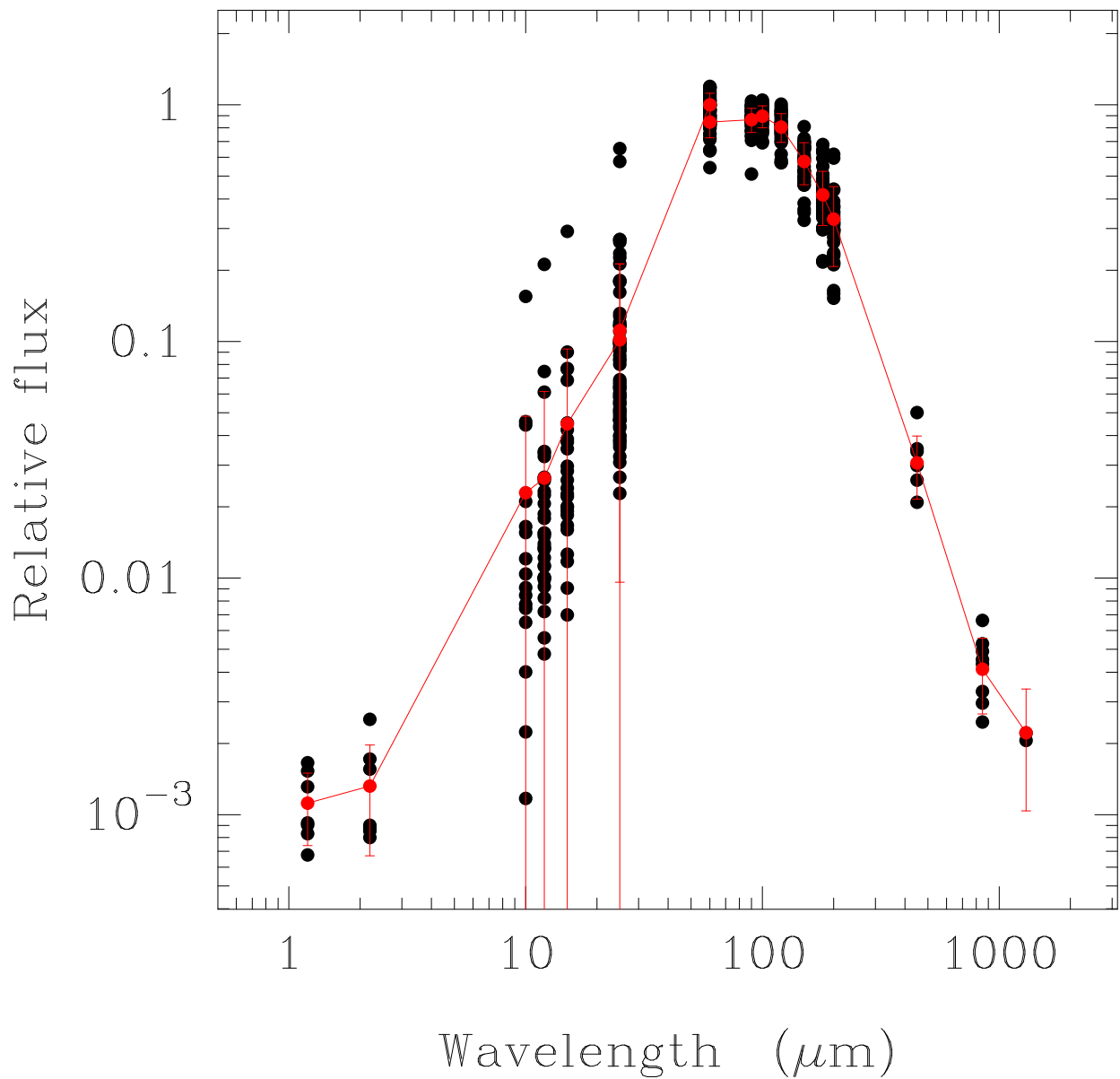}
\caption{The spectral energy distributions of the 37 ULIRGs part of
the local sample (from Klaas et al. 2001). Here the individual SEDs
have been multiplied by a constant factor, chosen such that the overall
dispersion is minimized. The individual data points, the average as well
as the 1$\sigma$ dispersions are shown (light grey). The line just connects
the average flux measurement at each observed wavelength.
\label{sedplot2}}
\end{figure}

In order to quantify the impression of a uniform shape of the SED, the
individual SEDs, $S_i$ where multiplied with a constant $A_i$ chosen
such that the overall dispersion is minimized.
In the present case, all the wavelengths were given equal weights.
No significant change could be found when restricting the wavelengths
to the FIR and sub-millimeter. The result is shown in
Fig.~\ref{sedplot2}, where the average flux (arbitrary normalization)
at each observed wavelength is shown together with the 1$\sigma$ deviation.
In Fig.~\ref{sedplot2} the resulting individual measurements are also
shown. It is clear that the impression of a general shape of the SEDs
is correct only for wavelengths $\lambda > 50\mu$m. At mid-IR wavelengths,
the dispersion is considerable. At near-IR wavelengths the dispersion is
again smaller than in the mid-IR.

\subsection{The method} \label{method}

\subsubsection{Fitting a single SED}
The small dispersion of the average spectral energy distribution of local
ULIRGs at wavelengths longer than 50$\mu$m (Fig.~\ref{sedplot2}) allows a
modified blackbody curve of the type discussed in Sect.~\ref{dustsed} to be
fitted to the average SED over this interval.
Although a two component fit is marginally superior to a single component,
especially around $90-120\mu$m, a single component dust SED was used for
simplicity.
The fitted function is given by Eq.~\ref{eq1} and the uncertainty associated
with each observed wavelength is taken to be the dispersion obtained from the
the average SED shown in Fig.~\ref{sedplot2}.

A least-square fit of Eq.~\ref{eq1} gives the following parameters:
$\Omega=(1.6 \pm 0.6) \times 10^{-13}$,  $\beta = 1.8 \pm 0.5$,
$\nu_0 = (1.2 \pm 0.4) \times 10^{12}$ Hz (250$\mu$m) and
$T_{\rm d} = 68 \pm 9$ K.
The $\Omega$ parameter is necessary for the fit but otherwise arbitrary
since the average SED has already been normalized. The $\chi^2$ is 0.89.
As emphasized in Sect.~\ref{dustsed}, the physical reality of the dust
properties associated with the fit is debatable. However, the shape of
the SED is well described by the fitted parameter.

\begin{figure}
\plotone{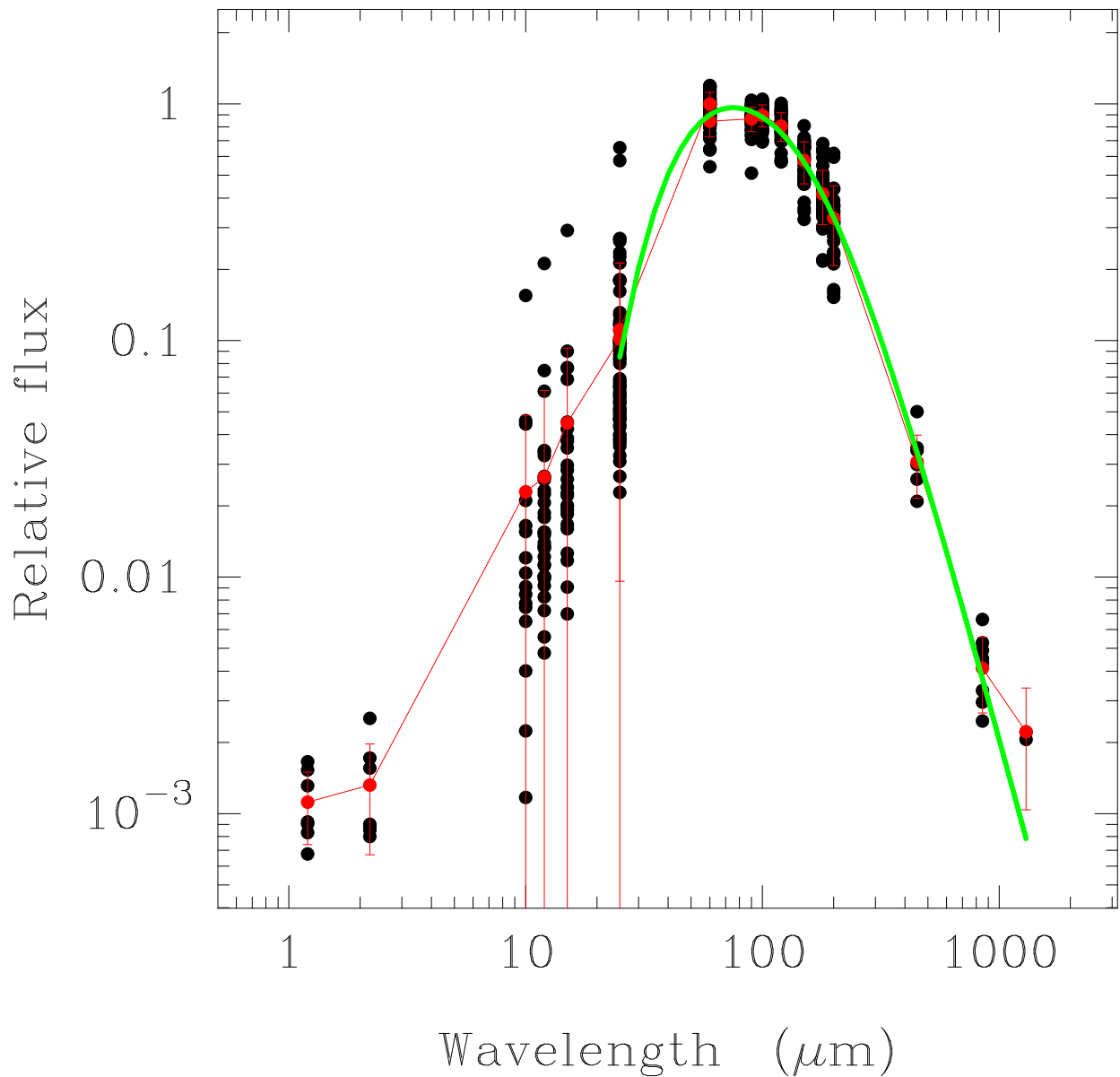}
\caption{A fit of a modified blackbody curve (see Eq.~\ref{eq1}) made
to the average SED obtained by minimizing the overall dispersion. The fit
has been made over the wavelength intervall 60-850$\mu$m. The best-fit
parameters are: $\Omega=(1.6 \pm 0.6) \times 10^{13}$, $\beta = 1.8 \pm 0.5$,
$\nu_0 = (1.2 \pm 0.4) \times 10^{12}$ Hz and $T_{\rm d} = 68 \pm 9$ K.
\label{sedfit1}}
\end{figure}

The resulting fitted SED is shown in Fig.~\ref{sedfit1} together with
the normalized data points. Only data in the wavelength range
$60 \leq \lambda \leq 850\mu$m were used in the fit.
The fitted curve coincides very tightly with all the average
data points in the range $25 \leq \lambda \leq 850\mu$m, but
only the $60-850\mu$m range will be considered in the following.
The 1250$\mu$m data were omitted because of a small number of
observations and the possibility of contribution from a very cold
dust component. Such a cold component would not be visible when
observing at shorter wavelengths and definitely not for sources
at high redshifts. The presence of such a cold dust component in
at least one of the sources in the present sample is suggested
by the excess flux density at $\lambda = 1250\mu$m seen in
Fig.~\ref{dustseds}.

\subsubsection{Monte Carlo simulation}
\begin{figure}
\plotone{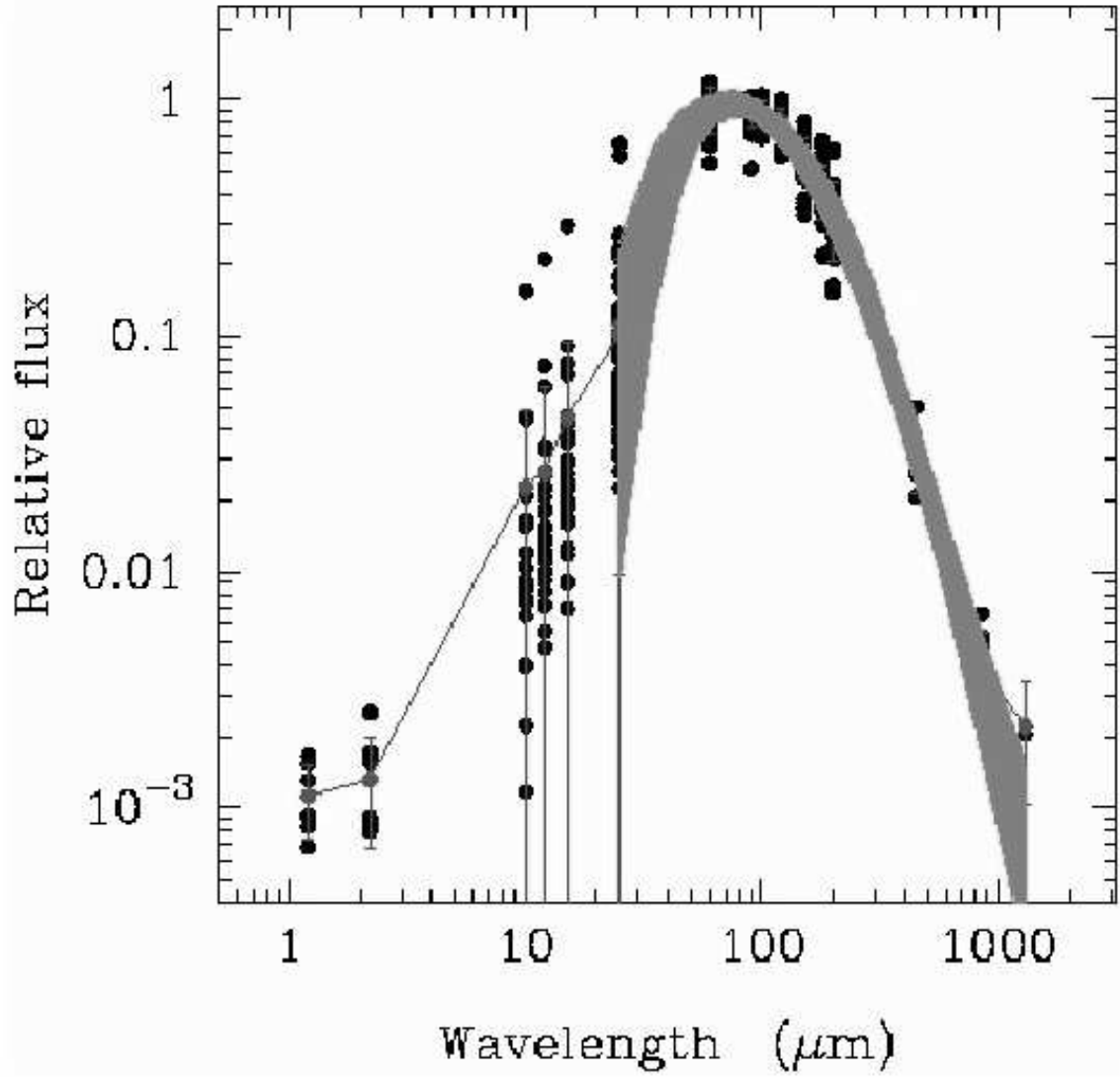}
\caption{Fits of modified blackbody curves resulting from letting the
flux density values at each wavelength vary uniformly $\pm1\sigma$. The
grey area represents $10^4$ simulated fits.
\label{sedfit2}}
\end{figure}

In order to estimate the uncertainty associated with the fit of
the single dust component, given the dispersion of the average
SED, a Monte Carlo simulation was done. For $10^4$ realizations the
average values of the underlying data points were stochastically
and uniformly varied within $\pm1\sigma$. For each simulated data
set a least-square fit of Eq.~\ref{eq1} was done. The fits were done
using data in the wavelength range $60\mu{\rm m} \leq \lambda
\leq 850\mu$m. The resulting SEDs are shown in Fig.~\ref{sedfit2}.
All the $10^4$ curves are plotted and they define an envelope which
remains well constrained in the wavelength range $60-850\mu$m, but
which diverges at shorter and longer wavelengths (where the fits
are no longer constrained by data). The fit to each of the Monte Carlo
realizations where done in the same manner as the fit discussed above.

\subsubsection{Flux ratios}
The observed flux at a given wavelength band is obtained by multiplying
Eq.~\ref{eq1}, using the fitted parameters, with a normalized filter
response function and then integrating over wavelength and normalizing
with the filter response function.
Here a simple top-hat function is used, with the same width as the filters
used on SCUBA, i.e. 30 GHz for both 450$\mu$m and 850$\mu$m. The same width
was also used for the 325$\mu$m band\footnote{This wavelength was chosen
in anticipation of the planned bolometer camera for APEX
(Atacama Pathfinder EXperiment).
The precise wavelength coverage and the shape of the filter
response function are not known presently. The results presented here for
this short wavelength band must therefore be recalculated once the filter
response is determined.}.
By redshifting the underlying SED and correcting for the $(1+z)$ compression,
the flux ratio at a given redshift was calculated.

\begin{figure}
\plottwo{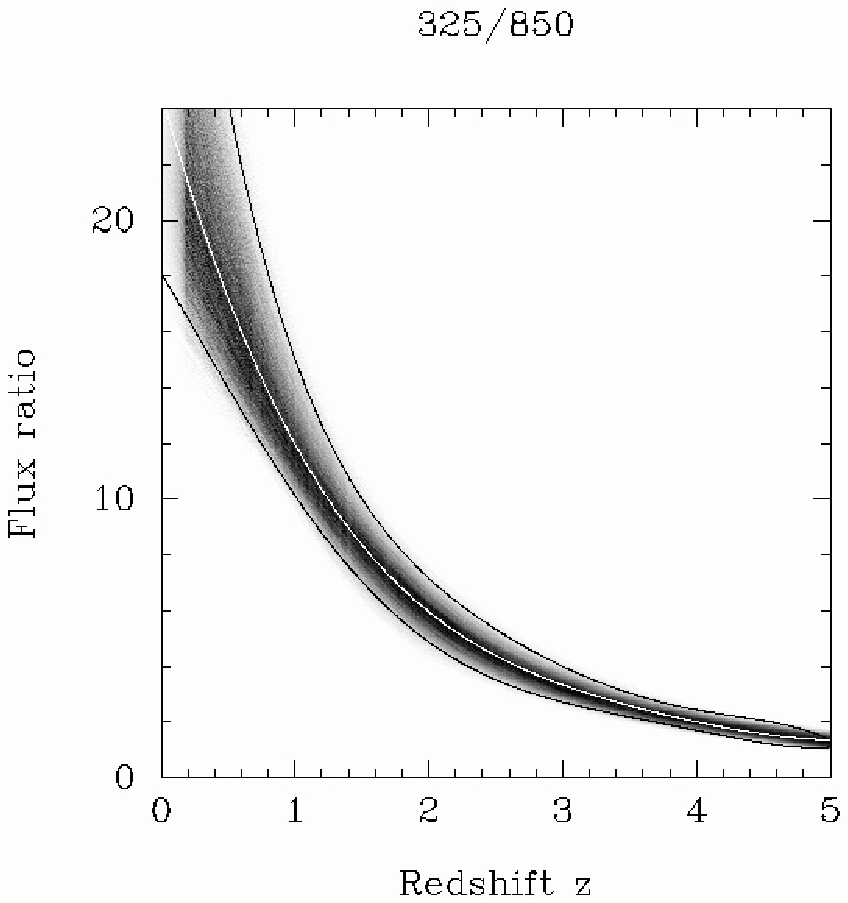}{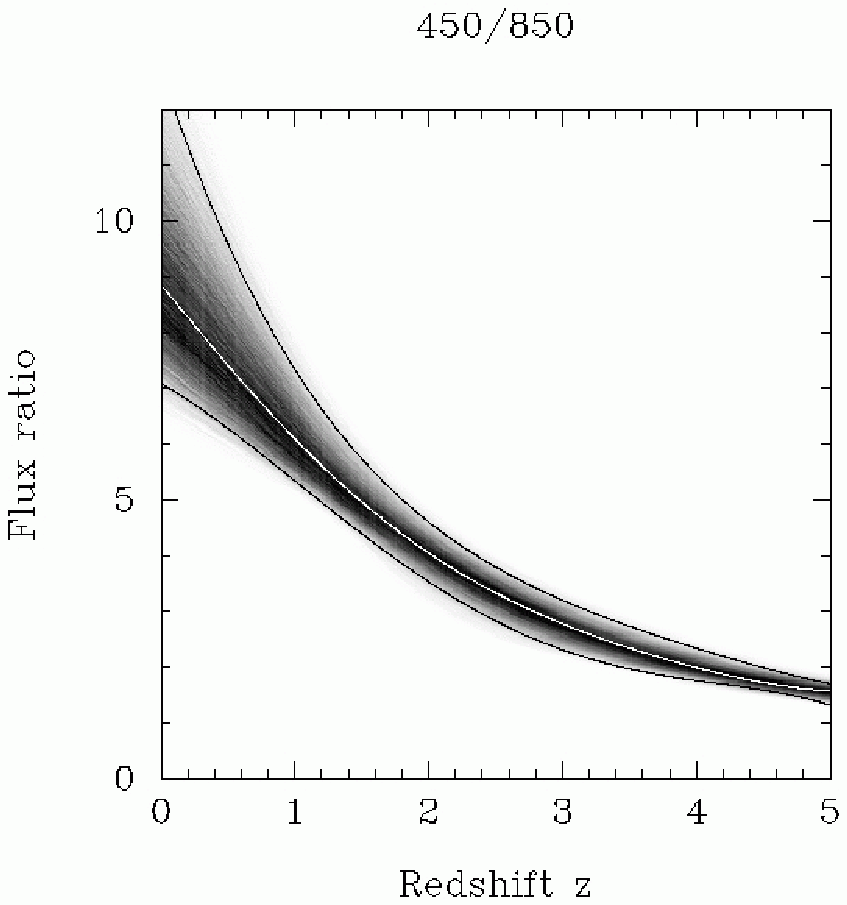}
\caption{The ratio of two fixed wavelength bands, each of width 30 GHz, as
a function of redshift of the emitting source.
The grey area corresponds to the results from $10^4$ Monte Carlo simulations
and represents the density of curves passing through a given point (dark means
high density). The maximum density is outlined by the white curve and it is
flanked by two curves representing the boundary containing $\pm95$\% of the
curves.
{\bf a)}\ The ratio of 325$\mu$m/850$\mu$m and {\bf b)}\ 450$\mu$m/850$\mu$m.
The 450$\mu$m and 850$\mu$m bands are chosen to represent the existing SCUBA
filters.
\label{greyplots}}
\end{figure}

The flux density ratio at two fixed wavelengths as a function of redshift
was obtained for each of the Monte Carlo realizations. They are plotted
in Fig.~\ref{greyplots} for both 325/850$\mu$m and 450/850$\mu$m.
The grey-scale corresponds to the density of curves passing
through a given point in the diagram, the darker the shade the higher
the density. The white full drawn line marks the ridge of highest density
and it is flanked by two lines marking the boundary containing 95\% of
the curves on either side of the ridge. By taking a cut at a constant
flux ratio, it is possible to assess the formal error in estimated redshift.
This error is for the most part non-gaussian.
In reality there will be an uncertainty associated with the measured
flux ratio as well. The predicted redshift will be enclosed by a box-shaped
region, limited by the high end of the flux ratio uncertainty at the lower
redshift end and the lower flux ratio uncertainty at the higher redshift
end.

At redshifts $z \leq 1$, the dispersion in redshift is too large for
the flux ratio to contain any predictive power of the redshift.
This is due to the fact that at small redshifts, both of the observed
wavelengths probe a section of the Rayleigh-Jeans part of the modified
blackbody curve with a close to a pure power law form. The flux ratio
therefore remains almost constant. The insensitivity to low redshifts
is less pronounced for the 325/850$\mu$m flux ratio since the shorter
wavelength will be affected by non-linearities in the dust SED at lower
redshifts than the 450$\mu$m band.
At high enough redshifts the shorter wavelength band will reach the Wien
side of the modified blackbody curve. The flux ratios will again start to
become more or less constant.
For the present sample this effect will set in at redshifts $z \geq 5$.
The most accurate predictive power for the redshift therefore exists in
the redshift range $1 \leq z \leq 5$.

\section{Discussion and Conclusion}

\subsection{Selection effects}

Any method of photometric redshift is based on the assumption that
we know or can model the spectral energy distribution of high redshift
objects. Modeling the FIR/submm SED is difficult due to a degeneracy
between the dust temperature and the grain emissivity power-law index
$\beta$. Moreover, as discussed in Sect.~\ref{degeneracy}, lowering
the dust temperature of a nearby source gives a similar effect on
the SED as moving the source to a higher redshift while keeping the
temperature constant. These degeneracies makes it impossible to use
the FIR/submm SED for photometric redshift determinations if one do not
have a sufficiently good apriori knowledge of the dust properties of 
the object in question.
There appears to exist a relation between the dust temperature and FIR
luminosity (e.g. Dunne et al. 2000; Blain et al. 2002), which can possibly
be used to break the degeneracy. This Luminosity-Temperature (LT) relation
is, however, presently poorly determined.

In Sect.~\ref{template} it was shown that the SED of local ultraluminous
IR galaxies has a remarkably similar shape at wavelengths longer than
$\lambda \ge 50\mu$m.
The similarity in the shape of the SEDs could very well be an effect of
sample selection. The local galaxies was selected as being ULIRGs, meaning
that they have a L$_{\rm FIR} \geq 10^{12}$ L$_{\odot}$. At the same time,
all the sources have L$_{\rm FIR} < 10^{13}$ L$_{\odot}$. If there exists
a LT-relation, this selection of sources within a limited luminosity range
favors those which have similar dust temperatures.
Whenever detailed properties are known for the high redshift submm detected
sources, which is the case for 2-3 objects, they have properties very
similar to the local ULIRGs in terms of L$_{\rm FIR}$, gas mass and star
formation. 
If the submm detected objects are at high redshift they must have L$_{\rm FIR}$
exceeding 10$^{12}$ L$_{\odot}$, or they would not be detected with present
day instrumentation. They are therefore likely to occupy the same region of
the LT-relation as the local ULIRGs.
Making the assumption that the high redshift population as a whole is similar
to the local ULIRGs, the local sample can be used as a template.

While the local sample shows a remarkably small dispersion in the shape of
the long wavelength part of their SEDs, they differ considerably at mid-IR
wavelengths. In this wavelength regime the IR emission is likely to show
optical depth effects. The resulting SED is then dependent on the
geometry of the emitting region and its orientation relative to the observer.
The presence of an AGN would also be most apparent in this wavelength regime
(see Klaas et al. 2001).
While the shape of the SED at mid-IR contains substantial diagnostic power
for the presence of a compact heating source for the dust, whether it is
massive star formation or an AGN, it does not affect the use of submm bands
for photometric redshifts as long as $z < 10$.

\subsection{Comparison with other methods}

The assumption made here of the existence of a template SED is exactly
the same as done when using the radio-FIR spectral index method (Carilli
\& Yun 1999, 2000). The main difference between the two method is that
the FIR SED method does not need to exclude radio-loud AGNs, and it
works out to redshifts $z\sim 5$. This is
important since it is very difficult or impossible to distinguish between
high redshift low-luminosity AGNs and non-AGNs among $\mu$Jy radio
continuum sources without milli-arcsecond imaging. One example of a
high redshift radio-loud AGN, detected both as a submm source and in
radio continuum is the gravitationally lensed H1413+117. The observed
submm flux is much lower than expected from the radio continuum flux
(Carilli \& Yun 1999). The sample used here consists of a mixture of
pure starburst and AGNs systems (Klaas et al. 2001); 8 are designated
as starburst systems, 13 as LINERs, 8 as Seyferts (2 Sy 1 and 6 Sy 2),
and 8 remain unassigned of a spectral class. As discussed above, the
presence of an AGN is most evident in the mid-IR range and not at
$\lambda > 60\mu$m.

An alternative photometric redshift method for use with FIR and submm
observations was presented by Hughes et al. (2002). This method use a
large number of template SEDs, containing both starburst systems and
AGNs, spanning a large range of intrinsic luminosities. The local IRAS
60$\mu$m  luminosity function is evolved with redshift to explain the
observed number-counts. The luminosity function is then randomly populated
by the SEDs from the template catalog.
Photometric redshifts are determined by calculating the probability
that the colors of an observed submm source are consistent with the colors
of every template SED at a given redshift. No correlation between the SED,
luminosity and redshift is assumed.
While models of this kind incorporate lower luminosity objects, the existing
instrumentation limits detection of high redshift objects to the most luminous
ones. Until submm observations can routinely be done with rms noise levels of
0.1\,mJy or lower, and in view of a possible luminosity-temperature relation,
it seems prudent to restrict the template to objects with luminosities large
enough to be detectable at high redshifts.

\subsection{Comparison with observations}

\begin{figure}
\epsscale{0.8}
\plotone{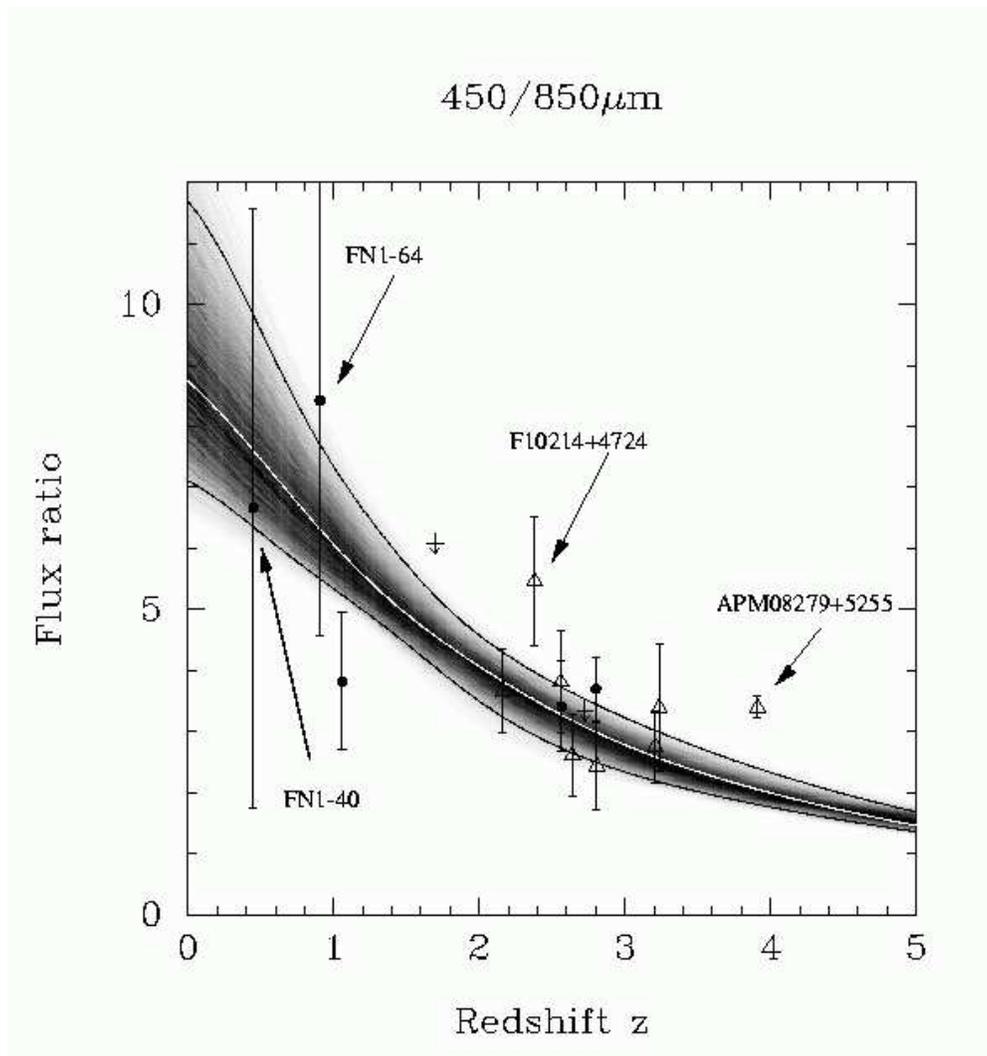}
\caption{The flux ratio of 450$\mu$m/850$\mu$m, measured at fixed wavelenghts,
versus redshift of the emitting source. The results from $10^4$ Monte
Carlo simulations are shown. The greyscale represents the density of curves
passing through a given point (dark means high density).
The maximum density is outlined by the white curve and it is flanked by
two curves representing the boundary containing $\pm95$\% of the curves.
Also shown are the existing sources observed at both 450$\mu$m and 850$\mu$m
and with known redshift. The open triangles are taken
from a sample of high redshift QSOs (Barvainis \& Ivison 2002). All of
these are gravitationally lensed but differential magnification is likely
to be significant only for the strongest lenses (IRAS F10214+4724 and
APM08279+5255). Filled circles correspond to submm detected galaxies.
\label{data450}}
\end{figure}

The flux density ratio 450$\mu$m/850$\mu$m, derived from the template,
as a function of redshift is shown in Fig.~\ref{data450}. Also shown
are measured flux density ratios for high-z sources with known spectroscopic
redshifts (see Table~\ref{table1}).
Six of these are taken from a survey of gravitationally lensed AGNs
(Barvainis \& Ivison 2002). Three are submm detected galaxies with known
redshift: SMM02399-0136 at $z=2.81$ (Ivison et al. 1998), SMM02399-0134
at $z=1.06$ and SMM1411+0252 at $z=2.56$ (Ivison et al. 2000). Two sources
detected by ISO at 170$\mu$m (Chapman et al. 2002):
FN1-40 at $z=0.45$ and FN1-64 at $z=0.91$. Two additional gravitational
lenses are also included, F10214+4724 at $z=2.3$ (Rowan-Robinson et al. 1993)
and APM08279+5255 at $z=3.91$ (Irwin et al. 1998).
The ISO sources are situated at relatively small redshifts and still
have large uncertainties associated with their observed 450$\mu$m flux
densities. Their predicted redshifts with the present photometric method
are therefore uncertain. Nevertheless, the present method achieves a
better result than the radio-FIR method which predicts redshifts of
1.1 and 1.3 instead of the spectroscopically derived 0.45 and 0.9.

Most of the sources shown in Fig.~\ref{data450} and listed in
Table~\ref{table1} are gravitationally lensed and it is possible
that differential magnification can distort the flux density ratios.
This can happen if the emission regions dominating at 450$\mu$m has
a smaller spatial extent than that of 850$\mu$m. The effect would be to
enhance the 450$\mu$m flux relative to that at 850$\mu$m and move the
data points upwards in the diagram. This, however, is only expected
to be of concern for the most strongly lensed sources.
Indeed, the data points deviating significantly from the expected
ratios are due to the strongest lenses (F10214+4724 and APM08279+5255).
Another deviant source is SMM\,J02399-0134 with a spectroscopic redshift of
$z=1.06$, where the indicated photometric redshift is $2.2^{+0.9}_{-0.6}$
(see Table~\ref{table1}). This, however, is a redshift where the present
method is less accurate due to the relatively small redshift.
Apart from these three sources, the agreement with the redshift expected from
the template model is good.

The low redshift sample of Dunne et al. (2001) generally fall below the
predicted flux density ratio (given their redshift). This sample, however,
only contains two sources which can be classified as ULIRG and hence, is not
expected to conform to the ULIRG template. In particular, their average
dust temperature is considerably lower (Dunne et al. 2001), in agreement
with a possible luminosity-temperature relation.

\begin{figure}
\epsscale{1.0}
\plotone{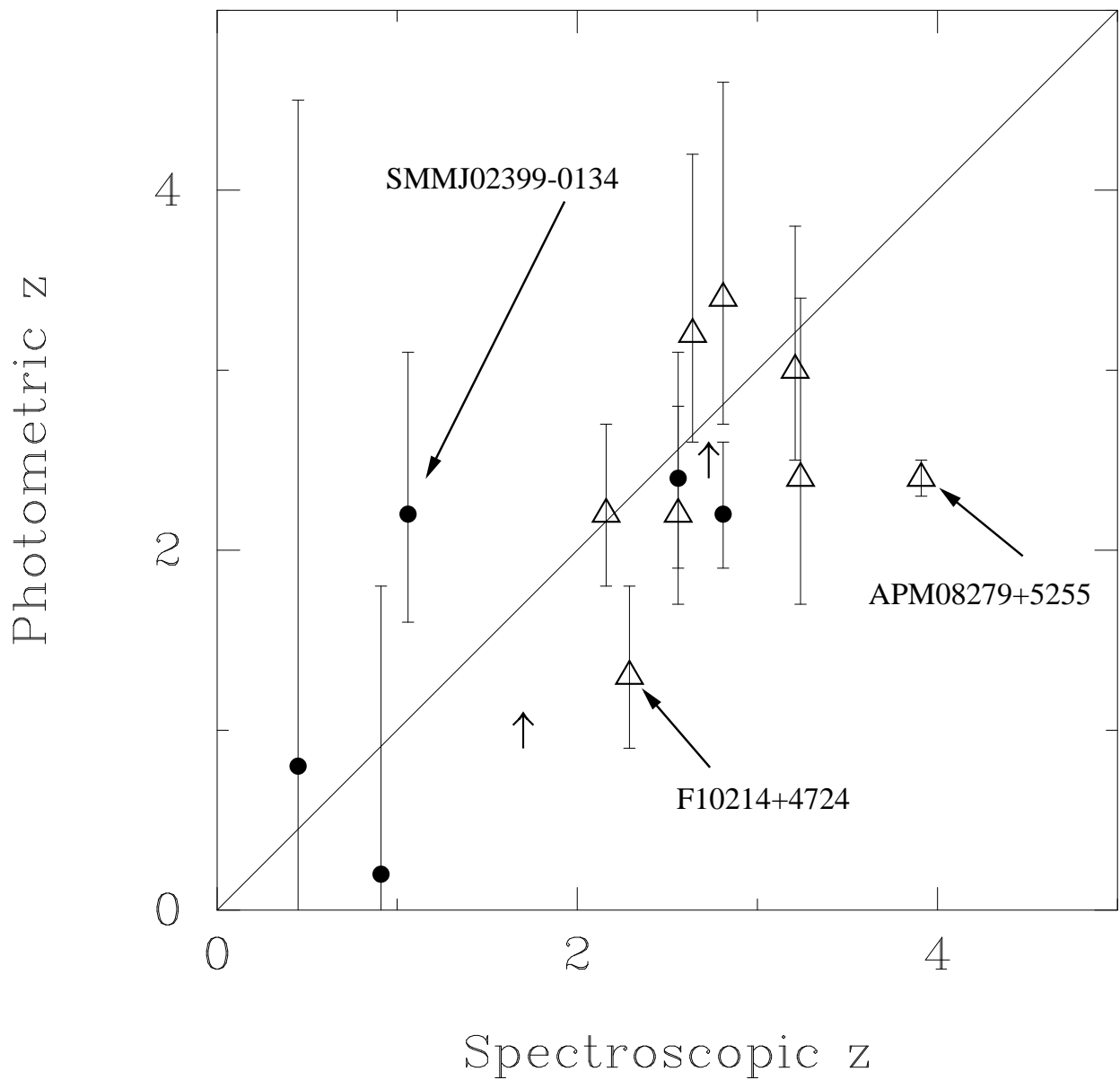}
\caption{Comparison between photometric redshifts using the present method
and spectroscopic redshift for a sample of submm detected objects and high-z
quasars (see Table~\ref{table1}). The full drawn line corresponds to a one-to-one
relation between photometric and spectroscopic redshifts and is not a fit to the
data. The errors in photometric redshifts have been derived using the observed
photometric errors convolved with the 95\% confidence level for the template.
\label{compfig}}
\end{figure}

The spectroscopic and photometric redshifts are given in Table~\ref{table1}.
The photometric redshifts are given with the errors calculated as a convolution
of the photometric uncertainty and the 95\% confidence limit derived from the
template SED. The spectroscopic and photometric redshifts are compared
graphically in Fig.~\ref{compfig}.
The uncertainty is dominated by the errors associated with the 450$\mu$m flux
densities rather than the spread in template flux ratios.

\subsection{Photometric to spectroscopic redshifts}

Ultimately it is preferable to determine the redshift of all the submm
detected galaxies spectroscopically. As mentioned in Sect.~\ref{intro}
this is in most cases very difficult due to the very weak emission at
optical and NIR wavelengths. An alternative method will become available
when ALMA (Atacama Large Millimeter Array) comes on-line. The sensitivity
of this interferometer will be high enough to allow detection of CO emission
from high redshift objects in a relatively short observing time (Blain et al.
2000). The instantaneous wavelength coverage will be much larger
than present day instruments. The projected bandwidth is 16 GHz, corresponding
to a redshift intervall $\Delta z/z \approx 0.2-0.3$ in the 3mm atmospheric
window. Although still relatively limited, it represents an order of magnitude
improvement to existing millimeter and submillimeter instruments.
The 3mm atmospheric window is ideal for searching for high redshift CO
emission. Except for a redshift intervall $0.44 \leq z \leq 1.0$, there
will always be at least one CO rotational transition within $80-115$ GHz.
For redshift $z > 3$ there will be at least two transitions within the
band.

\begin{figure}
\plotone{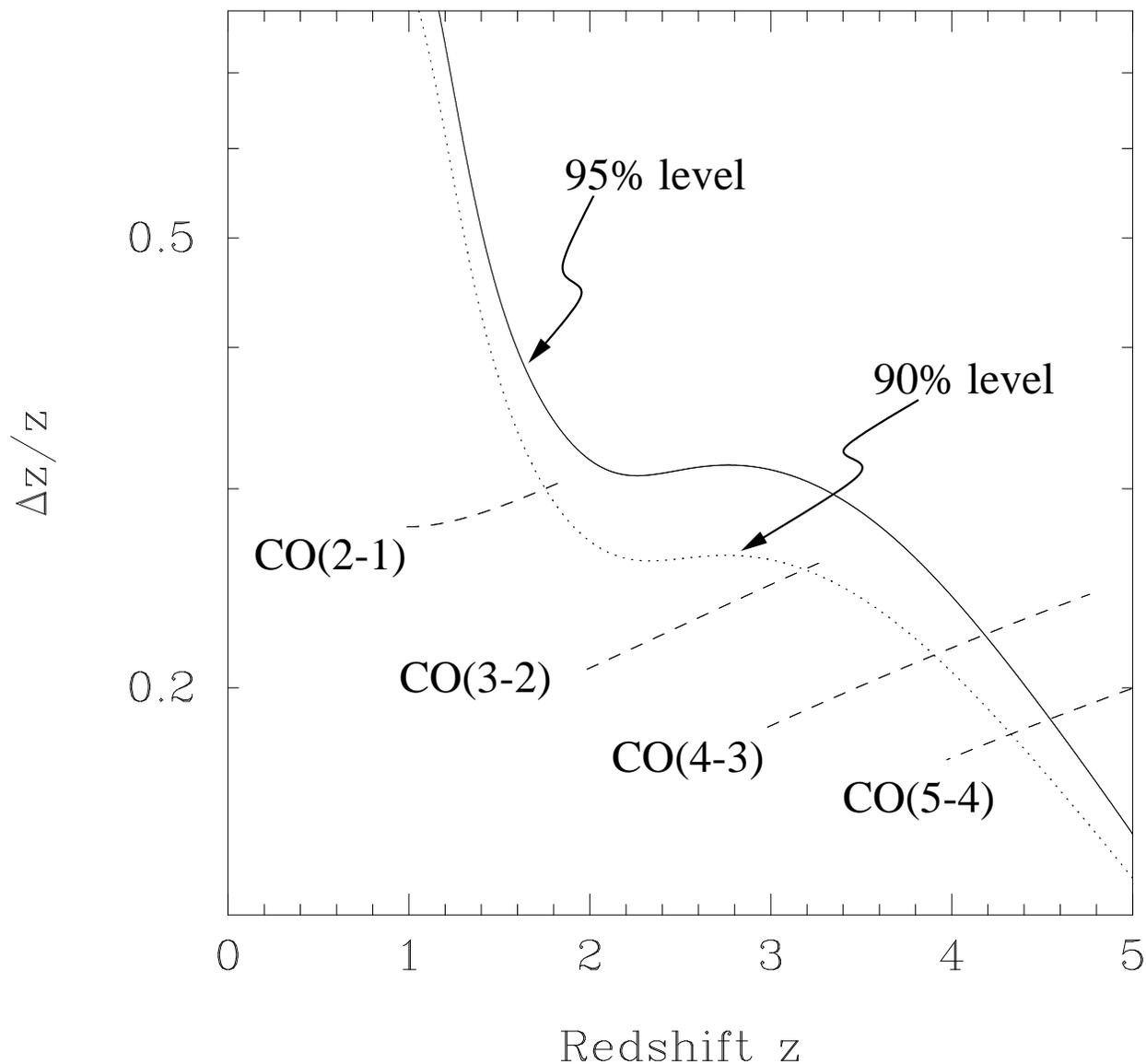}
\caption{The uncertainty in redshift associated with the template, using the
$\pm$95\% (full-drawn line) and $\pm$90\% (dotted line) confidence limits (the
$\pm$95\% confidence limit is shown in Fig.~\ref{data450}). While $\Delta$z/z
is large for small redshifts, it drops to $\sim$0.3 for $z>1.5$ and even further
at larger redshifts.
Also shown is the $\Delta z/z$ coverage of 4 CO rotational lines with the
planned ALMA receivers in the 3mm atmospheric windows. The redshift coverage
depends on the observed frequency (the bandwidth being fixed at 16 GHz) and
increases as a given transition is observed at higher redshift (lower 
observed frequency). Photometric redshifts are accurate enough to allow direct
tuning for transitions at approximately $z > 3$.
\label{deltafig}}
\end{figure}

If the photometric errors can be reduced to such a level that the uncertainty
in the photometric redshift is dominated by the dispersion in template SED,
the redshift uncertainty would be $\sim$30\% for redshifts $z > 1.5$.
This is illustrated in Fig.~\ref{deltafig}, where $\Delta z/z$ is plotted
as a function of redshift for the 95\% and 90\% confidence limit of the
template SED.
Also shown in Fig.~\ref{deltafig} are the corresponding $\Delta z/z$ values
of four CO transitions as observed in the 3mm band with the projected ALMA
receiver bandwidth of 16 GHz.
The 3mm band used here stretches from 80 GHz to 116 GHz, although
the first receiver implementation is likely to cover a somewhat smaller band.
Using higher frequencies results in a smaller redshift coverage and does not
cover the rotational transitions believed to give the strongest signal
(e.g. Combes, Maoli \& Omont 1999). For the highest redshifts covered by
the 3mm band, and thus the higher J-transitions, the redshift uncertainty
in the photometric method is comparable or smaller than the redshift
coverage of the ALMA bandwidth. For lower redshifts it will be necessary
to use two tunings in order to brackett a photometrically estimated
redshift. Improvements of the photometric redshift method may remedy this,
making it possible to use a single tuning of ALMA when searching for
redshifted CO emission even for lower redshift sources.

\section{Conclusion}

Despite a serious degeneracy between the dust temperature and the redshift
for a modified blackbody curve, it is shown that an average dust spectral
energy distribution, based on a local sample of ultraluminous IR galaxies,
can be used for photometric redshift determination in the intervall $1 < z < 5$. 

The local sample of ultraluminous IR galaxies (Klaas et al. 2001) shows a
remarkably small dispersion in the shape of the SED at wavelengths longward
of $\lambda = 50\mu$m. The local sample contains both starburst and AGN
dominated sources.

Using this template and the internal dispersion in its shape, it is shown that
observations at two fixed submm wavelengths, e.g. 450$\mu$m and 850$\mu$m, can
be used to determine the photometric redshift of galaxies in the redshift range
$1 < z < 5$. The uncertainty in the redshift arising from the template SED
amounts to $\Delta z/z \sim 0.3$, with an improvement at the highest redshift
range.

\bigskip

\noindent
{\bf Acknowledgement}\ 
Careful reading and valuable comments from Duilia de Mello is gratefully
acknowledged. Thanks to the referee U. Klaas for careful reading of the
manuscript and suggestions which improved the presentation of the paper.

\clearpage

\begin{deluxetable}{lrrc}
\tabletypesize{\small}
\tablecaption{SIMBA data at 1250$\mu$m \label{sesttable}}
\tablewidth{0pt}
\tablehead{
\colhead{Name}       & \colhead{S$_{1250}$}         & 
\colhead{$\sigma$}   & \colhead{Comments}           \\
\colhead{}           & \colhead{mJy}                &
\colhead{mJy}        & \colhead{}
}
\startdata
00199-7426      &    49 &   13 &           \\
05189-2524      &    11 &    3 &           \\
06035-7102      & $<$43 &   -- & 3$\sigma$ \\
06206-6315      & $<$13 &   -- & 3$\sigma$ \\
17208-0014      &    64 &   13 &           \\
20100-4156      &    21 &    6 &           \\
E148IG002       & $<$16 &   -- & 3$\sigma$ \\
23230-6726      & $<$21 &   -- & 3$\sigma$ \\
\enddata
\tablecomments{Sources from the Klaas et al. (2001)
sample observed with the SIMBA bolometer array at
the SEST in October 2002.
}
\end{deluxetable}

\clearpage

\begin{deluxetable}{lcrcrcccc}
\tabletypesize{\small}
\tablecaption{Submm flux data \label{table1}}
\tablewidth{0pt}
\tablehead{
\colhead{Name}       & \colhead{z$_{\rm spec}$}     & 
\colhead{$f_{850}$}  & \colhead{$\sigma_{850}$}     &
\colhead{$f_{450}$}  & \colhead{$\sigma_{450}$}     & 
\colhead{R\tablenotemark{a}}          & \colhead{z$_{\rm phot}$    } &
\colhead{Reference}  \\
\colhead{}           & \colhead{}                   &
\colhead{(mJy)}      & \colhead{(mJy)}              &
\colhead{(mJy)}      & \colhead{(mJy)}              &
\colhead{}           & \colhead{}                   &
\colhead{}
}
\startdata
SMM J14011+0252 & 2.56 & 12.3 & 1.7 &  42     &  7 & $3.4 \pm 0.7$ & $2.4^{+0.7}_{-0.5}$ & (1) (2) \\
SMM J02399-0136 & 2.81 & 23.0 & 1.9 &  85     & 10 & $3.7 \pm 0.5$ & $2.2^{+0.4}_{-0.3}$ & (1) (2) \\
HR10            & 1.44 &  8.7 & 1.6 &  $<$180 &    & $<$21         & --                  & (3) \\
SMM J02399-0134 & 1.06 & 11.0 & 1.9 &  42     & 10 & $3.8 \pm 1.1$ & $2.2^{+0.9}_{-0.6}$ & (1) \\
FN1-64          & 0.91 &  5.9 & 1.4 &  50     & 19 & $8.5 \pm 4.0$ & $0.2^{+1.6}_{-   }$ & (4) \\
FN1-40          & 0.45 &  6.3 & 1.4 &  42     & 30 & $6.7 \pm 5.0$ & $0.8^{+3.7}_{-   }$ & (4) \\
                &      &      &     &         &    &               & & \\
SMM J09429+4658 &  --  & 17.2 & 1.9 &  61     &  7 & $3.5 \pm 0.6$ & $2.3^{+0.4}_{-0.4}$ & (1) \\
SMM J14009+0252 &  --  & 15.6 & 1.9 &  33     &  9 & $2.1 \pm 0.6$ & $3.8^{+1.1}_{-0.8}$ & (1) \\
SMM J00266+1708 &  --  & 18.6 & 2.4 &  $<$60  &    & $<$3.2        & $>2.6$              & (5) \\
Lockman850.1    &  --  & 10.5 & 1.6 &  35     & 10 & $3.3 \pm 1.1$ & $2.5^{+1.1}_{-0.7}$ & (6) \\
HDF850.1        &  --  &  7.0 & 0.4 &  $<$21  &    & $<$3.0        & $>2.8$              & (7) \\
                &      &      &     &         &    &               & & \\
2237+0305       & 1.70 &  2.8 & 0.9 & $<$17   &    & $<$6.1        & $>1.0$              & (8) \\
HE0230-2130     & 2.16 & 21.0 & 1.7 &  77     & 13 & $3.7 \pm 0.7$ & $2.2^{+0.5}_{-0.4}$ & (8) \\
F10214+4724     & 2.29 & 50.0 & 5.0 & 273     & 45 & $5.5 \pm 1.1$ & $1.3^{+0.5}_{-0.4}$ & (8) \\
H1413+117       & 2.56 & 58.8 & 8.1 & 224     & 38 & $3.8 \pm 0.8$ & $2.2^{+0.6}_{-0.5}$ & (8) \\
MG J0414+0534   & 2.64 & 25.3 & 1.8 &  66     & 16 & $2.6 \pm 0.7$ & $3.2^{+1.0}_{-0.6}$ & (8) \\
UM673           & 2.73 & 12.0 & 2.2 &  $<$40  &    & $<$3.3        & $>2.5$              & (8) \\
RX J0911.4+0551 & 2.81 & 26.7 & 1.4 &  65     & 19 & $2.4 \pm 0.7$ & $3.4^{+1.2}_{-0.7}$ & (8) \\
MG J0751+2716   & 3.21 & 25.8 & 1.3 &  71     & 15 & $2.8 \pm 0.6$ & $3.0^{+0.8}_{-0.5}$ & (8) \\
CLASS B1359+154 & 3.24 & 11.5 & 1.9 &  39     & 10 & $3.4 \pm 1.0$ & $2.4^{+1.0}_{-0.7}$ & (8) \\
APM08279+5255   & 3.91 & 84.0 & 3.0 & 285     & 11 & $3.4 \pm 0.2$ & $2.4^{+0.2}_{-0.2}$ & (8) \\
\enddata

\tablenotetext{a}{$R = S(450\mu$m)/$S(850\mu$m)}

\tablecomments{References:
(1) Smail et al. (2002)
(2) Ivison et al. (2000);
(3) Cimatti et al. (1998);
(4) Chapman et al. (2002);
(5) Frayer et al. (2000);
(6) Lutz et al. (2001);
(7) Hughes et al. (1998);
(8) Barvainis \& Ivison (2002)}

\end{deluxetable}

\clearpage

\clearpage

\clearpage

\clearpage

\clearpage

\clearpage

\clearpage

\clearpage

\clearpage

\clearpage

\end{document}